\documentclass[aps,preprint]{revtex4}%
\usepackage{amsfonts}
\usepackage{amsmath}
\usepackage{amssymb}
\usepackage{graphicx}%
\begin{document}
\title[ ]{Any Classical Description of Nature Requires Classical Electromagnetic
Zero-Point Radiation}
\author{Timothy H. Boyer}
\affiliation{Department of Physics, City College of the City University of New York, New
York, New York 10031}

\begin{abstract}
Any attempt to describe nature within classical physics requires the presence
of Lorentz-invariant classical electromagnetic zero-point radiation so as to
account for the Casimir forces between parallel conducting plates at low
temperatures. \ However, this zero-point radiation also leads to classical
explanations for a number of phenomena which are usually regarded as requiring
quantum physics. \ Here we provide a cursory overview of the classical
electromagnetic theory which includes classical zero-point radiation, and we
note the areas of agreement and disagreement between the classical and quantum
theories, both of which contain Planck's constant $\hbar.$

\end{abstract}
\maketitle

\section{Introduction}

Although classical physics provides satisfactory explanations for many
phenomena in mechanics and electromagnetism, there seems to be little interest
in classical explanations for phenomena which involve Planck's constant
$\hbar$. Thus although there are natural classical explanations for the stable
ground state of hydrogen, for the blackbody radiation spectrum, for Casimir
forces, for specific heats of solids, and for diamagnetism, these explanations
are not mentioned in the physics textbooks. The root cause for this neglect is
the failure of modern physicists to allow the possibility of classical
electromagnetic zero-point radiation. In this article we start out by
discussing the experimentally observed Casimir forces where measurements have
become increasingly accurate in recent years. \ We note that these experiments
demand the presence of classical electromagnetic zero-point radiation if we
attempt to explain nature within classical electromagnetic theory. \ We then
point out that the presence of classical zero-point radiation has significant
implications for thermal behavior and atomic structure.

Some physicists will object that we already have perfectly good quantum
explanations for these phenomena so that classical explanations are
superfluous. \ To these physicists who like the quantum theory explanations,
we would simply repeat the words of Sherlock Holmes: "I don't mean to deny
that the evidence is in some ways very strong in favour of your theory; I only
wish to point out that there are other theories possible."\cite{ACD}

The classical theory which includes classical electromagnetic zero-point
radiation has in the past been termed "random electrodynamics"\cite{random} or
"stochastic electrodynamics"\cite{stochastic} or "classical electron theory
with classical electromagnetic zero-point radiation."\cite{with}\cite{DeLa}
\ The theory corresponds to the classical electron theory of H. A. Lorentz but
with a change in the boundary conditions to include classical electromagnetic
zero-point radiation. \ In recent years, the theory has had notable successes
in the simulation work for the hydrogen atom by Cole and Zou\cite{Cole} and
also in the relativistic work which provides an entirely new perspective on
blackbody radiation.\cite{submit} \ 

\section{Casimir Forces and Classical Electromagnetic Zero-Point Radiation}

\subsection{Casimir Forces}

The need for classical electromagnetic zero-point radiation within a classical
theory seems most transparent when we try to explain the experimentally
observed Casimir forces between conducting parallel plates. \ Casimir forces
are forces associated with the discrete normal mode structure of waves in a
finite volume.\cite{oned} \ Thus if we consider the thermal motion of a
one-dimensional string of length $L$ which has fixed end points at $x=0$ and
$x=L$, the random thermal motion can be expressed in terms of the oscillations
of the normal modes of the string with random phases between the modes.
\ Thermal wave motion will have a characteristic energy $U(\omega_{n})$
associated with each mode of (angular) frequency $\omega_{n}=2\pi nv/(2L),$
$n=1,2,...$ where $v$ is the speed of the waves on the string. \ If we imagine
the string passing through a small hole in a partition located at some point
$x$ between the fixed end points, $0<x<L,$ then the small hole will enforce a
node in the string's oscillations, and therefore the partition will experience
forces due to the oscillations of the string on the two different sides of the
partition. \ In general the partition will experience a net force because the
normal modes on opposite sides of the partition are associated with different
lengths $x$ and $L-x.$ \ This net force is a Casimir force on the partition
arising from the differences in energies $U(\omega_{n})$ for different
frequencies associated with the different lengths $x$ and $L-x$ of the string.
\ An analogous situation arises for any wave system where the boundary
conditions enforce a nodal structure. \ 

For a conducting partition in a conducting-walled box, electromagnetic waves
will lead to Casimir forces. The possibility of Casimir forces between
conductors was first proposed by H. B. G. Casimir\cite{CasimirF} in 1948 in
connection with the normal modes for electromagnetic radiation between
conducting parallel plates. \ Any spectrum of random radiation will lead to
forces on a conducting partition in a conducting-walled box. \ One of the
familiar spectra for random electromagnetic radiation is the Rayleigh-Jeans
spectrum where the energy $U_{RJ}(\omega,T)$ per normal mode at (angular)
frequency $\omega$ and temperature $T$ is independent of the frequency
$\omega$ and is given by $U_{RJ}(\omega,T)=k_{B}T$. \ This spectrum leads to
an attractive Casimir force $F_{RJ}$ between conducting parallel plates of
area $A=L\times L$ and separation $d,$ where $d<<L;$ the force is proportional
to the temperature $T$ and to the plate area $A=L\times L$, and inversely
proportional to the third power of the separation $d$ between the
plates\cite{PR1975}%
\begin{equation}
F_{RJ}=-\frac{\zeta(3)k_{B}TA}{4\pi d^{3}}\label{FF1}%
\end{equation}
where $\zeta(3)$ is a numerical constant. According to this formula, the
Casimir force should vanish as the temperature $T$ goes to zero. However,
experimental measurements\cite{Spaarnay} show clearly that the Casimir forces
do not vanish at low temperature, but rather become independent of
temperature. Within classical physics, the only natural explanation for the
experimentally measured Casimir forces between uncharged conducting plates at
low temperature is the existence of temperature-independent random radiation.
\ This radiation has been termed classical electromagnetic zero-point radiation.

\subsection{Spectrum of Classical Electromagnetic Zero-Point Radiation}

What is the natural spectrum for classical electromagnetic zero-point
radiation? This radiation should correspond to the state of lowest possible
energy, the vacuum state. And our qualitative notion is that the vacuum should
be as featureless as possible; in an inertial frame, it should be homogeneous,
isotropic, scale invariant, and indeed Lorentz invariant. It turns out that
there is a unique spectrum (unique up to one multiplicative constant) of
random classical radiation which satisfies these
requirements.\cite{stochastic}\cite{conf} The spectrum has an energy
$U_{0}(\omega)$ per normal mode given by
\begin{equation}
U_{0}(\omega)=\mathfrak{const}\times\omega/c \label{FF2}%
\end{equation}
where $\mathfrak{const}$ is an unknown constant. We mentioned that
\textit{any} spectrum of random classical radiation will lead to Casimir
forces between conducting parallel plates. The force between parallel
conducting plates of area $A=L\times L$ separated by a small distance $d$,
$d<<L,$ in the presence of the classical electromagnetic zero-point spectrum
of Eq. (\ref{FF2}) is given by
\begin{equation}
F_{0}=\mathfrak{const}\times\frac{\pi^{2}A}{120d^{4}} \label{FF3}%
\end{equation}
Indeed, it is found that this formula describes the experimental measurements
provided that
\begin{equation}
\mathfrak{const}=1.58\times10^{-26}J\cdot m \label{FF4}%
\end{equation}
Thus the experimentally observed Casimir forces at low temperature are
accounted for by the Lorentz-invariant spectrum of classical electromagnetic
zero-point radiation given in Eq. (\ref{FF2}) provided that the constant takes
the value in Eq. (\ref{FF4}).

\subsection{Planck's Constant $\hbar$ and Classical Electromagnetic Zero-Point
Radiation}

The constant appearing in Eq. (\ref{FF4}) was obtained from a purely classical
analysis of the Casimir forces between conducting parallel plates. \ No
aspects of energy or action quanta are involved. \ However, the numerical
value of the constant as well as the spectrum of Eq. (\ref{FF2}) are familiar
from a very different theory, namely from quantum theory. \ Thus we can either
continue to work with $\mathfrak{const}=1.58\times10^{-26}J\cdot m$ or we can
instead everywhere in the classical analysis replace $\mathfrak{const}$ by the
familiar expression $\hbar c/2$
\begin{equation}
\mathfrak{const}=\hbar c/2 \label{FF5}%
\end{equation}
since both have the same numerical value. \ 

Of course, there is a danger in introducing Planck's constant. \ Planck's
constant $\hbar$ has been associated with "quantum phenomena" for so long that
it is often referred to as "a quantum constant," and some physicists believe
that the mere presence of Planck's constant in a theory indicates that the
theory is a "quantum" theory. \ However, Planck's constant is simply a
numerical value which in itself does not indicate the type of theory where it
appears. \ As a numerical value, Planck's constant may appear in any theory.
\ Indeed, Planck's constant $h=2\pi\hbar$ was first introduced in 1899 before
there was any mention of quantum theory.

In the present discussion based upon classical electromagnetic theory,
Planck's constant $\hbar$ is introduced simply as a numerical value setting
the scale of classical electromagnetic zero-point radiation. \ As we have
emphasized above, we can avoid Planck's constant altogether simply by always
writing the expressions in terms of the $\mathfrak{const}$ appearing in Eq.
(\ref{FF4}) or by writing out its numerical value. \ The zero-point radiation
of Eq. (\ref{FF2}) is regarded as random classical radiation with a
Lorentz-invariant spectrum which appears as a homogeneous solution of
Maxwell's equations. \ Thus the solutions of Maxwell's equations in terms of
sources can be expressed as integrals over the sources using the retarded
Green function of the wave equation (thus providing the particular solution)
plus zero-point radiation as the homogeneous solution of Maxwell's equations. \ 

Perhaps the reader can obtain a sense of what is involved in classical
zero-point radiation by envisioning the more familiar situation involving
thermal radiation at nonzero temperature. \ Suppose that an experimenter sets
up his electromagnetic sources in a laboratory full of classical thermal
radiation at temperature $T>0.$ \ Then in order to describe the
electromagnetic fields in the lab, the experimenter would include both the
fields due to the sources which he has manipulated plus the fields due to the
thermal radiation which were already present and which the experimenter did
not introduce intentionally. \ And as every experimenter knows,
finite-temperature behavior will alter his sources. \ Thus the sources which
are introduced by the experimenter are influenced by the radiation which is
already present when the experimenter arrives in his lab. \ This radiation,
which is already present when the experimenter sets up his equipment,
corresponds to the homogeneous boundary condition on Maxwell's equations used
by the experimenter. Zero-point radiation is analogous to thermal radiation as
radiation which is not introduced by the experimenter but which is always
present and which can influence the sources which are arranged by an experimenter.

\section{Implications of Classical Electromagnetic Zero-Point Radiation}

\subsection{Linear Systems}

\subsubsection{Linear Oscillator}

Because classical electromagnetic zero-point radiation must be present in any
classical electromagnetic theory which accounts for the experimentally
observed Casimir forces between conductors, we also expect zero-point
radiation to influence every classical electromagnetic system. \ For example,
if we picture a particle of charge $e$ and mass $m$ at the end of a spring
oscillating along the $x$-axis so that the system has a natural mechanical
oscillation frequency $\omega_{0},$ then we expect that the system will both
be damped as the oscillating charge emits radiation and be pushed into motion
by the random zero-point radiation. \ Thus in the nonrelativistic point-dipole
approximation, we expect the system to satisfy the equation of motion%
\begin{equation}
m\frac{d^{2}x}{dt^{2}}=-m\omega_{0}^{2}x+\frac{2}{3}\frac{e^{2}}{c^{3}}%
\frac{d^{3}x}{dt^{3}}+e\mathbf{E}_{x}(0,t) \label{LL1}%
\end{equation}
where the mass times the acceleration equals the spring restoring force plus
the radiation damping force plus the zero-point radiation driving force.
\ This is a linear stochastic equation which can easily be solved. \ For a
small electric charge $e,$ the charge actually cancels out of the expressions
for the average values. \ The average position and momentum of the system are
both zero, but the mean square of the displacement $<x^{2}>$ and mean square
of the momentum $<p^{2}>=<(mv)^{2}>$ are given by\cite{with}\cite{general}%

\begin{equation}
<x^{2}>=\frac{\mathfrak{const}/c}{m\omega_{0}}=\frac{1}{2}\frac{\hbar}%
{m\omega_{0}} \label{LL2}%
\end{equation}
and
\begin{equation}
<p^{2}>=\frac{\mathfrak{const}}{c}\times m\omega_{0}=\frac{1}{2}\hbar
m\omega_{0} \label{LL3}%
\end{equation}
while the average energy is given by
\begin{equation}
U=\frac{\mathfrak{const}}{c}\times\omega_{0}=(1/2)\hbar\omega_{0} \label{LL4}%
\end{equation}
We have included the expressions involving $\mathfrak{const}$ so as to remind
the reader that these expression arise from the balance between the driving
force from classical zero-point radiation and the damping from the radiation
reaction force. \ However, it is clear that these expressions are identical
with those which appear in the quantum mechanics of the harmonic oscillator.
\ It turns out that the average values of all of the products of oscillator
position and momentum given by the classical calculations are identical with
the expectation values of the \textit{symmetrized} operator products of the
corresponding quantum oscillator.\cite{general}\ 

\subsubsection{Physical Systems Described by Linear Oscillators}

There are a number of physical systems which are traditionally described in
terms of molecules modeled as harmonic oscillators. \ These include the van
der Waals forces between molecules and also the van der Waals forces between
molecules and conducting or dielectric walls.\cite{van} \ The specific heats
of solids involve molecules which are often described by harmonic
oscillators.\cite{Santos} \ Finally, the diamagnetism of molecules can be
described in terms of the behavior of linear oscillator systems.\cite{dia}
\ Because of the general connection\cite{general} between the average values
of products of classical oscillator position and momentum with the symmetrized
operator products of the corresponding quantum variables, all of these
phenomena have natural classical descriptions within classical electromagnetic
theory which includes classical electromagnetic zero-point radiation.

\subsubsection{Disagreement for Nonrelativistic Nonlinear Non-Coulomb Systems}

One needs to be circumspect about the areas of agreement and disagreement
between classical and quantum theories. Despite the very close agreement for
linear systems between quantum theory and classical theory with classical
zero-point radiation, the theories part company for nonrelativistic nonlinear
non-Coulomb systems. \ Thus rotator specific heats are quite different within
quantum theory and classical theory with classical zero-point
radiation.\cite{rotator} \ Furthermore, classical nonlinear oscillator systems
scatter random radiation toward the Rayleigh-Jeans spectrum\cite{scatter}
whereas quantum systems do not.

\subsection{The Classical Hydrogen Atom}

One hundred years ago, Rutherford\cite{Ruth} published his work proposing the
nuclear model of the atom. \ Instead of the plum-pudding model for the atom,
consisting of a continuous "jelly" of positive charge with embedded negative
point electrons, the atom rather followed a "planetary" model, consisting of a
small, heavy, positive nucleus with electrons outside. \ However, it was
realized at the time that electrons in Coulomb orbit around the heavy nucleus
would radiate energy as electromagnetic radiation, and so it was thought that
they would spiral into the nucleus as they lost energy. \ At the time of
Rutherford's experiments, physicists were not aware of the idea of classical
electromagnetic zero-point radiation. \ The presence of this random zero-point
radiation, which we now know is required to exist in a classical theory so as
to account for Casimir forces, changes the perspective on the old problem of
atomic collapse. \ The presence of classical electromagnetic zero-point
radiation raises the possibility that atomic structure is due to a balance
between the loss of energy as electrons radiate and the pick-up of energy as
electrons experience the random forces of the zero-point radiation. \ This
basic model is the same as that used above in Eq. (\ref{LL1}) when discussing
linear oscillator systems. \ The nonrelativistic model for hydrogen
corresponds to the equation%
\begin{equation}
m\frac{d^{2}\mathbf{r}}{dt^{2}}=-\frac{e^{2}\widehat{\mathbf{r}}}{r^{2}}%
+\frac{2}{3}\frac{e^{2}}{c^{3}}\frac{d^{3}\mathbf{r}}{dt^{3}}+e\mathbf{E}(0,t)
\label{LL5}%
\end{equation}
where the mass times acceleration equals the Coulomb force attracting the
electron to the nucleus plus the radiation damping force plus the random force
due to the zero-point radiation. \ 

In contrast to Eq. (\ref{LL1}) for a linear system which was easy to solve
analytically, the stochastic differential equation (\ref{LL5}) for the
hydrogen atom has never been solved analytically. \ However, the ground state
has been solved by numerical simulation. \ In 2003, Cole and Zou\cite{Cole}
followed the motion of an electron described by equation (\ref{LL5}) and found
that the electron did not plunge into the nucleus or go far from the nucleus;
indeed, the probability distribution for the electron's distance from the
nucleus agreed closely with the familiar result given by the Schroedinger
ground state. \ There are no free parameters in Cole and Zou's calculation;
the values for the electron mass, charge, and scale of zero-point radiation
are all fixed by other experiments. \ The work is a striking suggestion of the
power of a classical theory in describing some parts of atomic physics.

There is also a revealing controversy associated with the calculation of the
hydrogen atom ground state. \ Cole and Zou's calculation provides a numerical
probability distribution for the hydrogen ground state and suggests that the
classical hydrogen atom with zero-point radiation is stable over the time.
\ On the other hand, Marshall and Claverie\cite{Claverie} in 1980 set up the
same nonrelativistic calculation in terms of action-angle variables. \ They
never computed any ground state distribution, but rather it was concluded that
there could be no stable ground state for the classical hydrogen atom in
classical electromagnetic zero-point radiation. \ The zero-point radiation was
viewed as "too strong," so that the electrons in Coulomb orbit around the
nucleus were ionized through the plunging orbits of small angular momentum.
\ Thus the work beginning with Marshall and Claverie's analysis suggested the
opposite situation from that of the old problem of atomic collapse in
classical theory. \ 

However, there is a failure in Marshall and Claverie's calculation as applied
to nature. \ Plunging elliptical orbits of small angular momentum do not exist
in nature! \ This situation often comes as a shock to physicists who are
familiar with the nonrelativistic classical mechanics of Coulomb and Kepler
orbits. \ Relativity changes the orbits of mechanical motion most severely for
orbits of small angular momentum.\cite{relorbit} \ Within the relativistic
mechanics of a point mass held in a Coulomb or Kepler orbit by a force
$\mathbf{F=-}$ $e^{2}\widehat{\mathbf{r}}/r^{2}$, any orbit which has small
angular momentum, $L<e^{2}/c,$ must plunge into the nucleus while conserving
energy and angular momentum! \ It should be emphasized that this last sentence
involves pure relativistic mechanics, not electromagnetism, and there is no
energy loss or gain due to radiation emission or absorption.\cite{relorbit}

Thus we find that the calculations of Marshall and Claverie are modified by
relativity at precisely the point where they suggest that the electron is
ejected from the atom. \ We conclude that Cole and Zou's numerical simulations
have indeed found the nonrelativistic approximation to the ground state of the
classical hydrogen atom in classical electromagnetic zero-point
radiation.\cite{comments}

\subsection{Blackbody Radiation and Relativity}

The beginning of the twentieth century saw the introduction of quantum ideas
in connection with the problem of blackbody radiation. \ Indeed today, those
textbooks which still introduce quanta from a historical\cite{Eisberg} rather
than an axiomatic\cite{Griffiths} perspective still discuss the classical
physics of radiation normal modes and energy equipartition within
nonrelativistic statistical mechanics.\cite{Eisberg} \ The blackbody radiation
problem troubled physicists all though the first quarter of the century. \ In
addition to the now-famous quantum calculations, there were attempts to derive
the equilibrium spectrum of thermal radiation from classical scattering
calculations\cite{scatter} and from equilibrium classical particle
motion.\cite{EH} \ However, the physicist in the first quarter of the
twentieth century were unaware of two important aspects of classical physics:
classical electromagnetic zero-point radiation and the importance of
relativity. \ 

The mere presence of classical electromagnetic zero-point radiation alters our
ideas of classical statistical mechanics. \ Indeed a number of derivations of
the Planck spectrum for blackbody radiation have been given within classical
physics based upon the presence of classical electromagnetic zero-point
radiation, some using as their starting point precisely the earlier
calculations which (in the absence of classical zero-point radiation) led to
the Rayleigh-Jeans spectrum.\cite{black} \ However, all of those calculations
left a nagging doubt because of the scattering calculations using
nonrelativistic charged mechanical systems; all of these calculations show
that nonrelativistic nonlinear scattering systems push classical radiation
toward the Rayleigh-Jeans spectrum.\cite{scatter} \ 

The importance of relativity appeared above in validating the work of Cole and
Zhou against the conclusion of Claverie and Marshall. \ The importance of
relativity also appears in understanding the spectrum of blackbody radiation.
\ Only in 2010 was it pointed out that a relativistic scattering system will
not scatter classical electromagnetic zero-point radiation toward the
Rayleigh-Jeans spectrum.\cite{relscat} \ This result is absolutely crucial.
\ All of the calculations leading to the Rayleigh-Jeans spectrum for classical
thermal radiation involve mixtures of nonrelativistic physics and relativistic
electromagnetic radiation.\cite{scatter} \ None of the calculations leading to
the Rayleigh-Jeans law holds up as a fully relativistic
calculation.\cite{Blanco}\ 

Indeed most recently, it has been shown that the Planck spectrum for thermal
radiation follows from the presence of classical zero-point radiation and the
structure of relativistic spacetime within classical physics.\cite{submit}
\ Zero-point radiation is the unique spectrum of random classical radiation
which is Lorentz invariant and scale invariant. \ Zero-point radiation is
required in the classical theory so as to account for the experimentally
observed Casimir forces. \ Now classical electromagnetism is invariant under
not only relativistic transformations but also conformal transformations which
include dilatations and proper conformal transformations.\cite{BC1909} \ In an
inertial frame, classical thermal radiation is carried into thermal radiation
at a different temperature by a time-dilating conformal transformation, while
the spectrum of classical zero-point radiation is invariant under
time-dilating conformal transformations and indeed under all conformal
transformations. \ However, if we consider thermal radiation not just in an
inertial frame but in a general non-inertial, static coordinate frame, then
time-dilating conformal transformations carry zero-point radiation into
thermal radiation at finite non-zero-temperature. \ We can use this connection
between zero-point radiation and thermal radiation in a coordinate frame
undergoing uniform relativistic acceleration through flat spacetime (a Rindler
frame\cite{Rindler}) to give a derivation of the Planck spectrum in an
inertial frame by taking the zero-acceleration limit.\cite{submit} \ The
Planck spectrum is connected directly with zero-point radiation and relativity
in classical physics.

\subsection{Speculations Regarding Wave-Like Aspects of Particles and Line
Spectra}

Although classical physics can give satisfying classical explanations for some
phenomena involving Planck's constant $\hbar$, there are at present no
calculations which give a definite explanation for the experimentally observed
diffraction effects for particles passing through small slits. \ Clearly
classical electromagnetic zero-point radiation interacts with any conducting
or dielectric surface, and this interaction changes the correlation function
for the electromagnetic fields compared to free space. \ Also, qualitatively,
the motion of charged particles near slits will be influenced by the
correlation function for the classical electromagnetic zero-point radiation
near the slits. \ Perhaps one day we will be able to calculate this influence
within a classical theory. \ In any case, it is obvious that if the number of
slits is changed, then the correlation function for zero-point radiation will
change, and hence the pattern of particles passing through the slits will
change. \ Indeed, the influence of surfaces which change the correlation
functions for zero-point radiation is well-understood for charged harmonic
oscillator systems at rest outside plane surfaces where the changes in the
correlation function lead to van der Waals forces.\cite{van}

Furthermore, there is at present no definitive explanation for the
experimentally observed line spectra. \ Within classical physics, we expect
line spectra to correspond to some sort of resonance behavior. \ In the
hydrogen atom, a highly excited electron radiates away more energy than it
picks up from the zero-point radiation, and indeed the spectra of hydrogen do
indeed approach the traditional spectral frequencies calculated in the absence
of zero-point radiation. \ This is the idea which is involved in the
correspondence principle. \ However, when the electron is excited but near the
ground state in energy, it seems hard to calculated the radiation emitted by
the electron as it loses energy by radiation emission and absorbs energy from
the classical electromagnetic zero-point radiation. \ Cole and Zou have
pointed out that the Coulomb potential holds fascinating nonlinear resonances
for electrons in a circularly polarized electromagnetic driving
wave.\cite{Cole2} \ However, there is at present no explanation within
classical phyisics for the line spectra of atoms.

\section{Discussion}

In this article, we have pointed out that any attempt at a classical
explanation of nature must included classical electromagnetic zero-point
radiation to account for the experimentally observed Casimir forces between
conducting parallel plates. \ The spectrum of classical zero-point radiation
can be determined up to one multiplicative constant by symmetry requirements,
such as Lorentz invariance or scale invariance or conformal invariance. \ The
scale of the classical zero-point radiation is determined by fitting
Casimir-force experiments. \ The numerical value obtained for the scale of
zero-point radiation is familiarly given in terms of the number appearing in
Planck's constant $\hbar$. \ Thus within classical theory, Planck's constant
$\hbar$ enters the theory as a number setting the scale of the homogeneous
solution of Maxwell's equations corresponding to classical electromagnetic
zero-point radiation. \ 

Once classical electromagnetic zero-point radiation is introduced into the
classical theory, there are implications for van der Waals forces, specific
heats, diamagnetism, atomic structure, and blackbody radiation. \ In some
cases, the classical calculations are in agreement with quantum theoretical
calculations, and in some cases they are not. \ For mechanical systems
described in terms of free fields or linear oscillator systems, there is
agreement between the classical and quantum average values. \ For
nonrelativistic nonlinear non-Coulomb systems, there is disagreement between
the classical and the quantum calculations. \ For the ground state of
hydrogen, there is fascinating agreement between classical numerical
simulation calculations and the Schroedinger ground state. \ Also, classical
physics gives a simple and powerful explanation for the blackbody radiation
spectrum in terms of zero-point radiation and relativistic theory.

Planck's constant $\hbar$ is a number which can appear in classical or quantum
theories. \ The constant $\hbar$ appears in all theories which include
zero-point radiation or zero-point energy. Within quantum theory, Planck's
constant is related to commutators of operators which then lead to zero-point
energy for quantized mechanical systems and zero-point radiation for quantized
fields. Within classical electron theory with classical electromagnetic
zero-point radiation (stochastic electrodynamics), Planck's constant appears
as a scale factor for classical zero-point radiation, and Planck's constant
then reappears in all systems with electromagnetic interactions. Thus Planck's
constant $\hbar$ is introduced at very different points in quantum as compared
to classical theory. What should we say about the limit $\hbar\rightarrow0$
which is often called "the classical limit"? \ The limit $\hbar\rightarrow0$
turns the quantum theory of noncommuting operators into a classical theory
with commuting variables, but any idea of zero-point energy or zero-point
radiation has disappeared along with quantum operator behavior. On the other
hand, the limit $\hbar\rightarrow0$ removes the zero-point energy from any
classical theory, and therefore this limit turns the classical electron theory
with classical electromagnetic zero-point radiation into the classical
electron theory of H. A. Lorentz\cite{Lorentz1952} which was used in the years
around 1900. Clearly the more recent classical electron theory which includes
classical zero-point radiation with a scale set by $\hbar$ can explain far
more of nature than is possible with the older classical electron theory where
zero-point radiation is omitted and $\hbar$ is regarded as zero.

\end{document}